\documentclass[fleqn,usenatbib]{mnras}

\usepackage[T1]{fontenc}
\usepackage{ae,aecompl}

\usepackage{graphicx}	
\usepackage{amsmath}	
\usepackage{amssymb}	

\makeatletter
\renewcommand\paragraph{\@startsection{paragraph}{4}{\z@}%
            {-2.5ex\@plus -1ex \@minus -.25ex}%
            {1.25ex \@plus .25ex}%
            {\normalfont\normalsize\it}}
\makeatother
\title[PAH in Wolf-Rayet Winds]{Search for Polycyclic Aromatic Hydrocarbons in the Outflows from  Dust-Producing Wolf-Rayet Stars}

\author[S. Marchenko and A. Moffat]{
Sergey V. Marchenko$^{1}$\thanks{E-mail: sergey.marchenko@ssaihq.com (SVM)}
and A. F. J. Moffat$^{2}$
\\
$^{1}$Science Systems and Applications, Inc., 10210 Greenbelt Rd., Lanham, Maryland, 20706, USA\\
$^{2}$D\'epartement de Physique and CRAQ, Universit\'e de Montr\'eal, C.P. 6128, Succ. Centre-Ville, Montr\'eal, Qu\'ebec, H3C 3J7, Canada
}

\date{Accepted XXX. Received YYY; in original form ZZZ}

\pubyear{2016}

\begin{document}
\label{firstpage}
\pagerange{\pageref{firstpage}--\pageref{lastpage}}
\maketitle

\begin{abstract}
A combined mid-IR spectrum of five colliding-wind, massive, dust-producing Population I Wolf-Rayet (WR) binaries shows a wealth of absorption and emission details coming from the circumstellar dust envelopes, as well as from the interstellar medium. 
The prominent absorption features may arise from a mix of interstellar carbonaceous grains formed in high- (e.g., 3.4, 6.8, 7.2 $\micron\,$) and low-temperature (3.3, 6.9, 9.3 $\micron\,$) environments. The broad emission complexes around $\sim$6.5, 8.0 and 8.8 $\micron\,$  could arise from ionized, small polycyclic aromatic hydrocarbon (PAH) clusters and/or amorphous carbonaceous grains. As such these PAH emissions may represent the long sought after precursors of amorphous Carbon dust. We also detect a strong $\sim$10.0 $\micron\,$ emission in the spectra of WR48a and WR112, that we tentatively link to ionized PAHs. Upon examining the available archival spectra of prodigious individual WR dust sources, we notice a surprising lack of 7.7 $\micron\,$ PAH band in the spectrum of the binary WR19, in contrast to the apparent strength of the 11.2, 12.7 and 16.4 $\micron\,$ PAH features. Strong PAH emissions are also detected in the $\lambda>$10 $\micron\,$ spectrum of another dust-producing system, WR118, pointing to the presence of large, neutral, presumably interstellar PAH molecules towards WR19 and WR118.  
\end{abstract}

\begin{keywords}
infrared: stars -- stars: winds, outflows -- stars: Wolf-Rayet -- (ISM:) dust, extinction
\end{keywords}



\section{Introduction}

The broad, strong emission bands observed in the mid-IR spectra of various astrophysical sources are 
commonly linked to partially hydrogenated, either positively charged or neutral, polycyclic aromatic hydrocarbon  molecules 
(L\'eger and Puget 1984; Sellgren, K. 1984; Allamandola et al. 1985). PAH features exibit a wide variety of spectral shapes and strengths (Peeters et al. 2002, van Diedenhoven et al. 2004), 
being sensitive to the hardness of the radiation field (e.g., Draine \& Li 2001, Hunt et al. 2010), and the chemistry (e.g., Ciesla et al. 2014) and density of the ambient environment.

In a modern-epoch galaxy roughly half of the PAH emission may come from the surroundings of evolved stars: mainly from asymptotic giant-branch stars (Galliano et al. 2008a), but also from planetary nebulae (Gillett et al. 1973), massive red giants (Melbourne \& Boyer 2013), dust-producing red super-giants (RSG: Verhoelst et al. 2009) and Luminous Blue Variables (LBV: Skinner 1997; Voors et al. 1999; Voors et al. 2000). The remaining PAH emissivity is produced by the radiation fields of ionizing sources (e.g., Lu et al. 2014), though 
the contribution from different sources is clearly site- and wavelength-dependent (Melbourne and Boyer 2013). PAH bands are also recorded in the UV-irradiated, photo-dissociated environments of massive stars
(Bern\'e et al. 2013) and PAH emissions are ubiquitous in young stellar object sites (Oliveira et al. 2013). In starburst galaxies PAH emission scales with star formation rate 
(D\'iaz-Santos et al. 2008), often dominating mid-IR spectra of H II regions. 
PAH emissions are detected even in the hostile environments of active galactic nuclei (AGN: Shipley et al. 2013; Esquej et al. 2014). Moreover, the PAH species seemingly survive in the immediate ($<$10 pc) nuclear vicinity (Alonso-Herrero et al. 2014), weakened  
due to dilution by the ANG emission rather than direct PAH destruction (Ramos Almeida et al. 2014). Indeed, the detailed mapping of PAH sizes in the 
reflection nebula NGC 7023 (Croiset et al. 2016) points to the possibility of a net growth of PAH clusters in the presence of strong UV sources.
Massive, luminous, He-burning Population-I Wolf-Rayet stars are especially interesting in this context, since they are both evolved and exceptionally strong UV-ionizing sources, closely adhering to active, star-forming regions.

Population I WR stars come in two groups: classical He-rich He-burning pre core-collapse SN Ib,c stars of initial mass $\ga25 M_\odot$ with subtypes WN, WC and WO, and main-sequence H-rich WNh stars of initial mass $\ga75 M_\odot$. Although numerous different evolutionary scenarios have been proposed for WR stars, that of Crowther (2007) is probably preferred:

1. $M_i\ga75 M_\odot$: [O] $\rightarrow$ WN(H-rich) $\rightarrow$ LBV $\rightarrow$  WN(H-poor) $\rightarrow$ WC $\rightarrow$ SNIc

2. $M_i \sim 40-75 M_\odot$: O $\rightarrow$ LBV $\rightarrow$ WN(H-poor) $\rightarrow$ WC $\rightarrow$ SNIc

3. $M_i \sim 25-40 M_\odot$: O $\rightarrow$ LBV/[RSG] $\rightarrow$ WN(H-poor) $\rightarrow$ SNIb

Square brackets show our modifications that are likely to apply: excluding an initial O stage for WN(H-rich) stars and adding a RSG stage for the lower masses in the 25-40 $M_\odot$ range. Among all the WR paths, only Carbon-rich WC stars sometimes exhibit IR emission excesses that reflect the formation of hot, Carbon-based dust.  Among these, a large fraction of the coolest subtype WC9 (and some WC8) stars, as well as a few WC+O binaries with any WC subclass, are often copious dust producers. Among the WC9/8 `dustars', it remains uncertain whether the colliding-wind compression in a binary is needed to make the hot dust, unlike the few WC+O binary dustars, where this does appear to be an integral part of the dust-formation process. Currently, the binary status of WC9 stars is being actively examined (Williams et al. 2005).

Note that, overall, Hydrogen is substantially (WN class) or completely (WC class) depleted in evolved WR stars. The lack of Hydrogen undermines the possibility 
of PAH formation via traditional pathways (J\'ager et al. 2009, Cherchneff 2010). However, large-scale mixing (Pittard 2007) of the Hydrogen-rich 
stellar wind of the relatively unevolved, early-type companion with the Carbon-rich wind of a WC star in a colliding-wind massive binary 
offers a potential way around the problem of Hydrogen deficiency, thus leading to the formation of the PAH species (Cherchneff 2011,2015), once we set aside the fundamental problem of PAH and dust formation in the extremely hostile environments of the hot ($T_{eff} \ga$40000 K) stellar winds. Here $T_{eff}$ denotes the effective temperature of the inner wind of WR stars at the optical depth $\tau = 2/3$, well above the hydrostatic core, often estimated to correspond to the level with $\tau=20$. 

The near- to mid-IR SEDs of continuously dust-producing WC stars are usually devoid of any spectral features, save for presumably interstellar absorptions (van der Hucht et al. 1996, Schutte et al. 1998, McCall et al. 2002). Most of the objects reviewed in this paper, except WR 19, belong to this category of continuous dust emitters. The second broad category, WR 19 inclusive, shows periodic dust outbursts that are seen as a relatively rapid growth and eventual, more gradual fading of a practically featureless mid-IR continuum emission from a hot ($T\ga$1000 K) dust presumably formed in the colliding-wind interface. 

So far, the reports on the WR-PAH features were sparse and sometimes controversial. 
Cohen et al. (1989) observed $\sim$8.0 $\micron\,$ emission profiles in the spectra of WR104 and WR112. Williams et al. (1994) identified a weak, broad 8.6 $\micron\,$ emission in the spectrum of the colliding-wind binary WR125 and tentatively related it to PAHs. Chiar and Tielens (2001) provided a tentative link of the 6.2 and 6.5-8 $\micron\,$ absorptions to amorphous Carbon and PAHs, respectively. The first unequivocal detection of PAH emissions at 6.4 and 7.9 $\micron\,$ in the dust-producing Wolf-Rayet star WR48a followed shortly (Chiar et al. 2002b).  

In order to verify these  claims, we decided to re-analyse the ISO-SWS IR spectra (van der Hucht et al. 1996) of the five most prodigious WR dust producers. We complement the repeatedly studied ISO-SWS data with excerpts from the abundant SPITZER library and use previously unpublished Gemini spectroscopy in the 7.5-13.5 $\micron\,$ domain. We show that massive colliding-wind WR binaries provide ample evidence of PAHs presumably formed in their circumstellar environments. 
We show that the list of detectable PAH features, either of circumstellar or interstellar origin, spans a broad range from 3 $\micron\,$ to at least 16.5 $\micron\,$. 

There is unquestionable proof of copious amounts of dust generated in the winds of the binary systems discussed in this paper. However, understanding of the underlying physics is fragmentary at best. The possible production of PAH in these winds only augments this formidable theoretical challenge. There are two conceivable ways of PAH 
production in the outflows from evolved stars (Cherchneff 2011). The 
PAH species may form via (presumably kinetic) chemical routes, thus 
preceding the formation of amorphous carbon dust. Or, alternatively, 
PAH may come as by-products of dust sputtering. Indeed, there is a 
growing consensus that PAH species may survive in the harshest 
environments, including the proximities of AGNs. Unfortunately, the evidence
collected in this study 
is too scant to allow us to conclude on the preferable pathway of PAH formation in WR colliding winds.

If some of the detected PAH features are indeed formed in WR winds, then in the early Universe such species could be produced by the first generations of massive stars, well before the traditional sources of PAHs take over the scene.  Hence, current-epoch WR-PAH spectra may provide a proxy to the properties of primordial cosmic dust.

\section{Observations}

In principal, each WR `dustar' source could provide somewhat different PAH spectrum. However, here we neglect such deviations in order to enhance S/N and make the PAH features more visible. The combined WR-PAH spectrum examined in this work comprises the individual mid-resolution ISO-SWS spectra of five WR stars described by van der Hucht et al. (1996), namely WR48a, WR98a, WR104, WR112 and WR118 (following the designations from van der Hucht 2001), all known as 
prodigious dust producers (e.g., Williams 2008). These spectra were extensively examined by various researchers (e.g., Schutte et al. 1998, Chiar et al. 2002).  

 \begin{figure*}
 	\includegraphics[width=170mm]{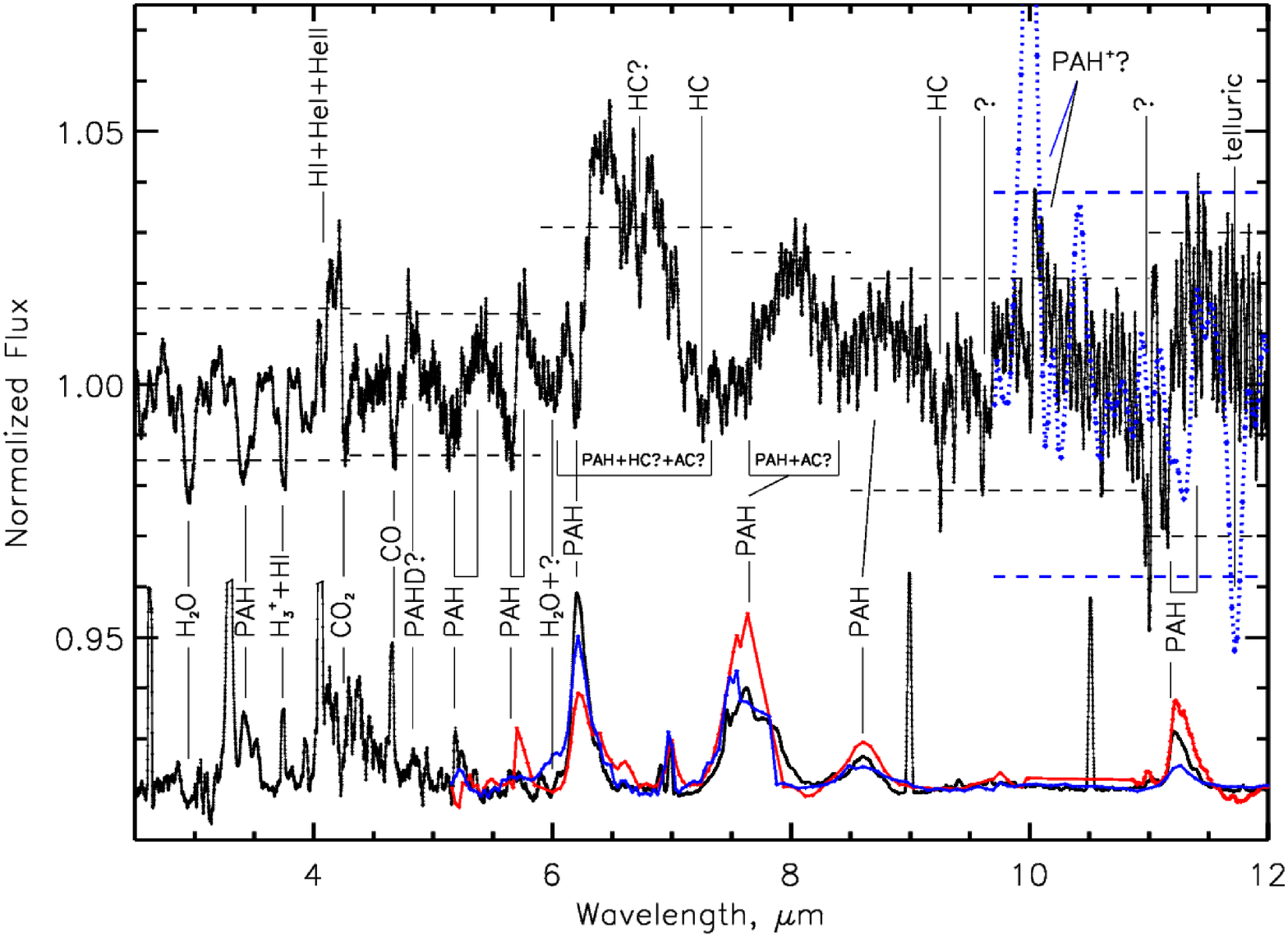}   
   \caption{The composite WR-PAH spectrum. Upper section: black line follows the rectified, normalized, then combined ISO-SWS spectrum of 5 Galactic WR `pinwheels', with the dashed lines defining 
the $\pm1\,\sigma$ detectability limits; blue dotted/dashed lines follow the combined, rectified Gemini-South spectrum of WR48a and WR112 with the $\pm1\,\sigma$ detection limits, respectively. 
For illustrational purposes, the lower section shows the arbitrarily scaled spectra of the Orion Bar (black), the Galactic-Center ISM (red) and the combined spectrum (blue) of the Galactic-Center stars WR102a (WN8) and WR102b (WN6), shifted by -0.08 (for clarity).}
    \label{fig1}
\end{figure*}

\begin{figure}
 	\includegraphics[width=\columnwidth]{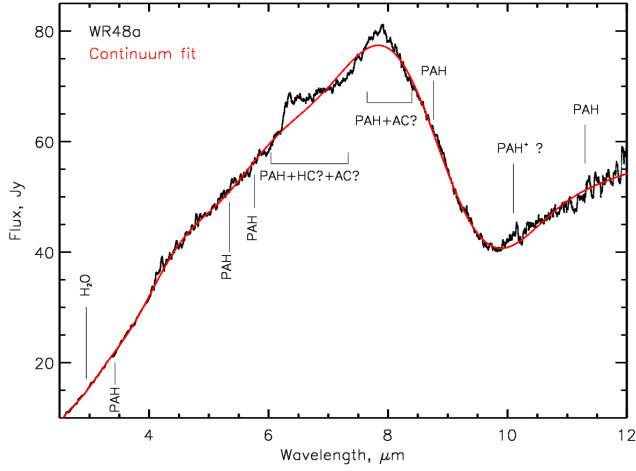}      
   \caption{The smoothed ISO-SWS spectrum of WR48a with continuum fit. }
    \label{fig2}
\end{figure}

The construction of the combined WR-PAH spectrum (Fig.~\ref{fig1}) required rectification of the individual WR spectra prior to final combination 
and smoothing, thus compromising the direct use of PAHFIT (Smith et al. 2007) for more elaborate decomposition of the PAH components.
In order to produce individual rectified spectra, for each of the five dust-making WR binaries we fitted the original ISO-SWS data with high-order (up to 25th: Figs.~\ref{fig2}-6) polynomials, normalized the original spectra by these fits and smoothed the spectral ratios by a running-mean (n=51 spectral points) filter. Inspecting the individual fits, one may evaluate the extent of uncertainties brought by the polynomial approximation. The 6-8 $\micron\,$ region carries relatively strong emission 
features that are clearly detected ($\sim$6.5 $\micron\,$ in particular) in all analyzed stars. However, the deep, broad 9.7 $\micron\,$ IS absorption borders the region of interest, making the fitting extremely challenging. This leads to $\sim$50\% overestimate of the 6-7$\micron\,$ emission complex in WR118, counter-balanced by a similar-magnitude under-estimate in WR112 (Figs.~\ref{fig5}-~\ref{fig7},~\ref{fig9}). Such 
occasional polynomial mis-fits may alter the strengths of individual PAH 
features in the  6-8$\micron\,$ range. However, for all targets 
except WR118, the polynomial fits turn out to be 
smooth enough for adequate reproduction of the shapes of PAH features, 
especially in the $\lambda <$6 $\micron\,$ region populated by 
relatively narrow lines. 

To produce a final WR-PAH composite spectrum, we opted for a direct, unweighted average of the individual spectra of all five WR stars (Figs.~\ref{fig7}-9), in order to enhance the detection of relatively weak, although prevailing spectral features. 
We identified the spectral features in the combined WR-PAH spectrum by inspecting the individual 
WR spectra. Then, governed by the preliminary S/N estimates we singled out the spectral details seen in at least 3 of the 5 spectra comprising the WR-PAH composite. 
Whenever appropriate, we fitted the selected features with single or multiple Gaussian functions. Table 1 lists the main properties of the 
identified spectral features: central wavelength, equivalent width, full-width at half-maximum, and the potential main carrier. In Figs.~\ref{fig1},~\ref{fig7}-~\ref{fig9} we also show provisional, rather conservative $\pm\,1\sigma$ detection limits. For each given wavelength range, these were derived as an average of the individual rms in the normalized, rectified spectra, assuming that they are inherently featureless. We noticed a rapid decline of the S/N ratio longward of 10 $\micron\,$, hence we limit our analysis to the $\lambda<$12 $\micron\,$ region. This particular limit was chosen to include the diagnostically useful (see below) 11.3 $\micron\,$ PAH feature. For qualitative comparisons we also use one of the archived Galactic Center spectra taken by Spitzer during Cycle 1 (Simpson et al. 2007), as well as the spectra from the Spitzer Atlas (Ardila et al. 2010). These spectra were fitted, then normalized by polynomials of 4-6 order, smoothed by a running-mean filter, then arbitrarily (for comparison purposes) scaled.

\begin{figure}
 	\includegraphics[width=\columnwidth]{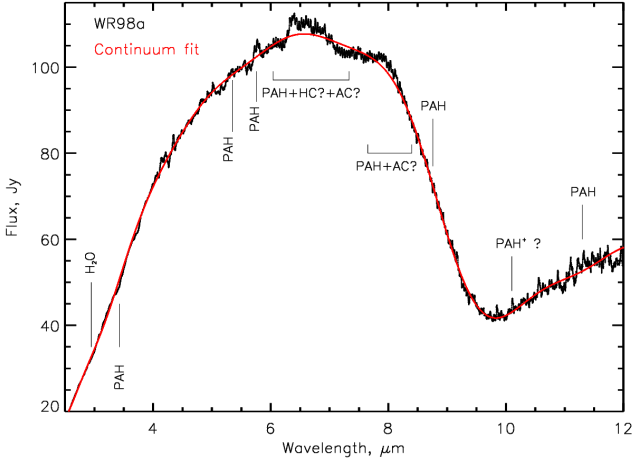}      
   \caption{Same as in Fig.~\ref{fig2}, but for WR98a. }
    \label{fig3}
\end{figure}

\begin{figure}
 	\includegraphics[width=\columnwidth]{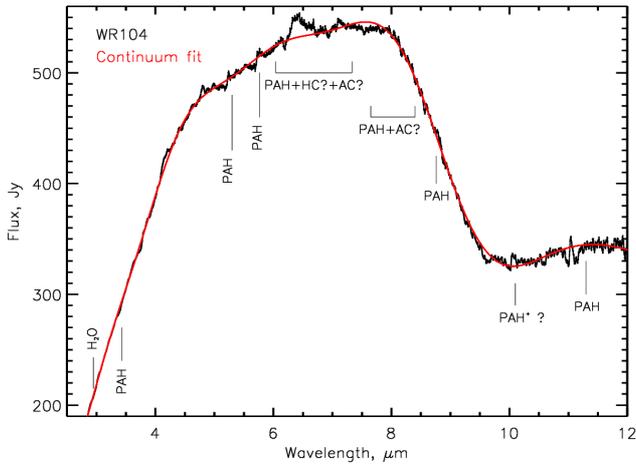}      
   \caption{Same as in Fig.~\ref{fig2}, but for WR104. }
    \label{fig4}
\end{figure}

\begin{figure}
 	\includegraphics[width=\columnwidth]{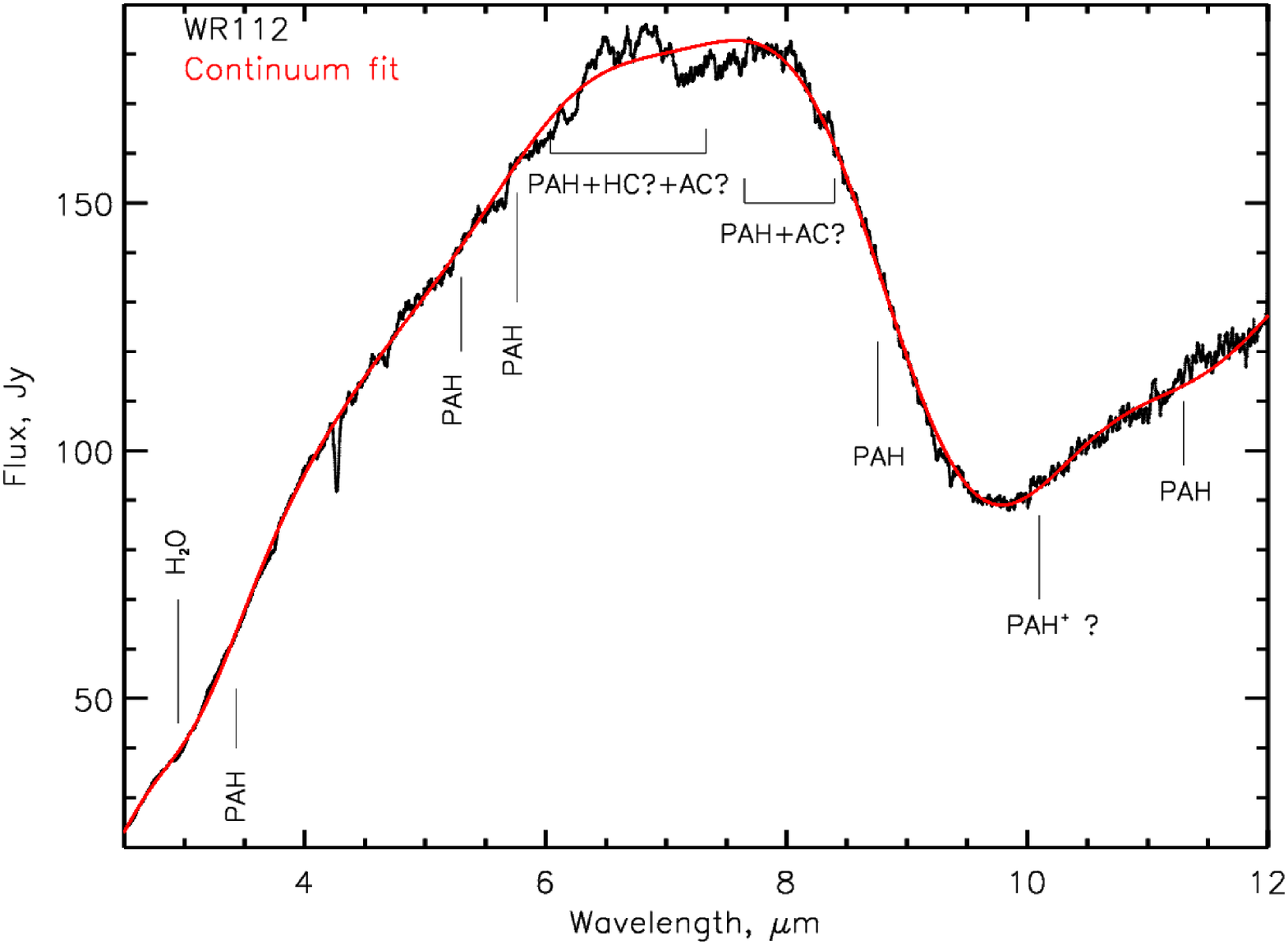}      
   \caption{Same as in Fig.~\ref{fig2}, but for WR112. }
    \label{fig5}
\end{figure}

\begin{figure}
 	\includegraphics[width=\columnwidth]{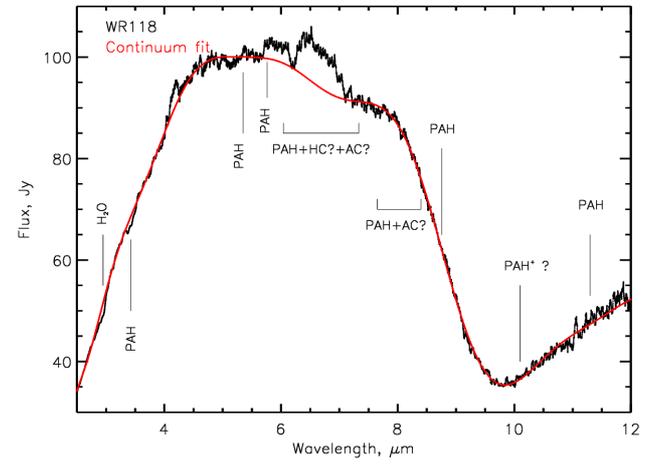}      
   \caption{Same as in Fig.~\ref{fig2}, but for WR118. }
    \label{fig6}
\end{figure}

On April 12,17 and May 15, 2006 two WR `dustars', WR48a and WR112, were observed with the Thermal-Region Camera and Spectrograph
(T-ReCS: De Buizer and Fisher 2005) mounted on the 8-meter Gemini-South telescope. The observations yielded multiple exposures, each with accumulated 10-20 min on-source time. The long-slit, low-resolution ($R\sim$100) 7.5-13.5 $\micron\,$ T-ReCS spectra were processed with IRAF \footnote {IRAF is distributed by the National Optical Astronomy Observatories, which are operated by the Association of Universities for Research in Astronomy, Inc., under cooperative agreement with the National Science Foundation.} routines by appropriately combining all the available data from the chopping and nodding modes of observations. The contemporaneous (same-night) observations of IR standards, HR 715 and HR 4617,  were used for reduction of the telluric spectral features. Further processing comprised polynomial fitting and normalization of individual exposures for both WR objects, then their combination into an average, normalized spectrum shown in Fig.~\ref{fig1}. 

\begin{figure*}
 	\includegraphics[width=16cm]{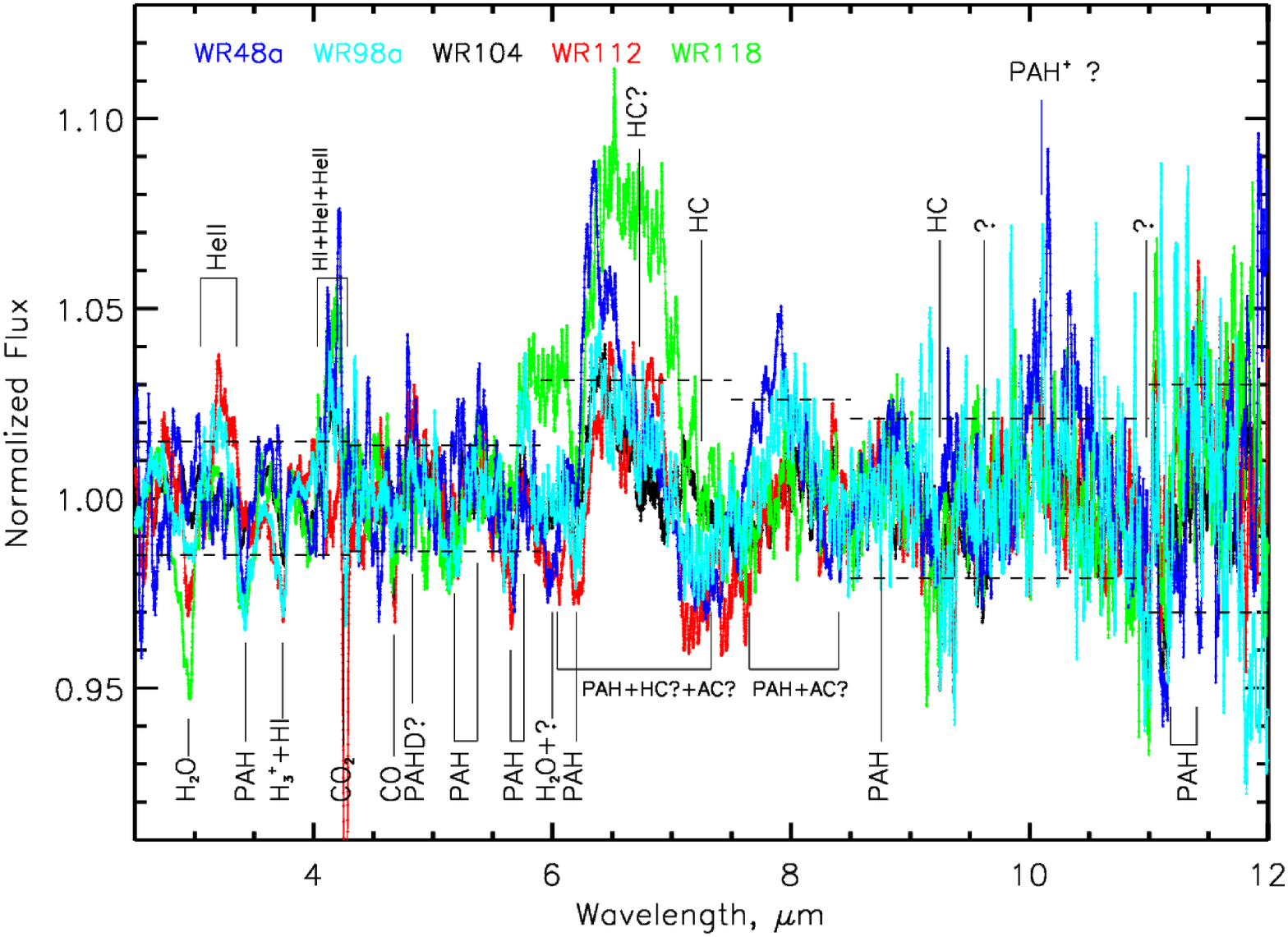}      
   \caption{Individual continuum-normalized ISO-SWS spectra of 5 WR `pinwheels' contributing to the combined WR spectrum in Fig.~\ref{fig1}. }
    \label{fig7}
\end{figure*}

\begin{figure}
 	\includegraphics[width=\columnwidth]{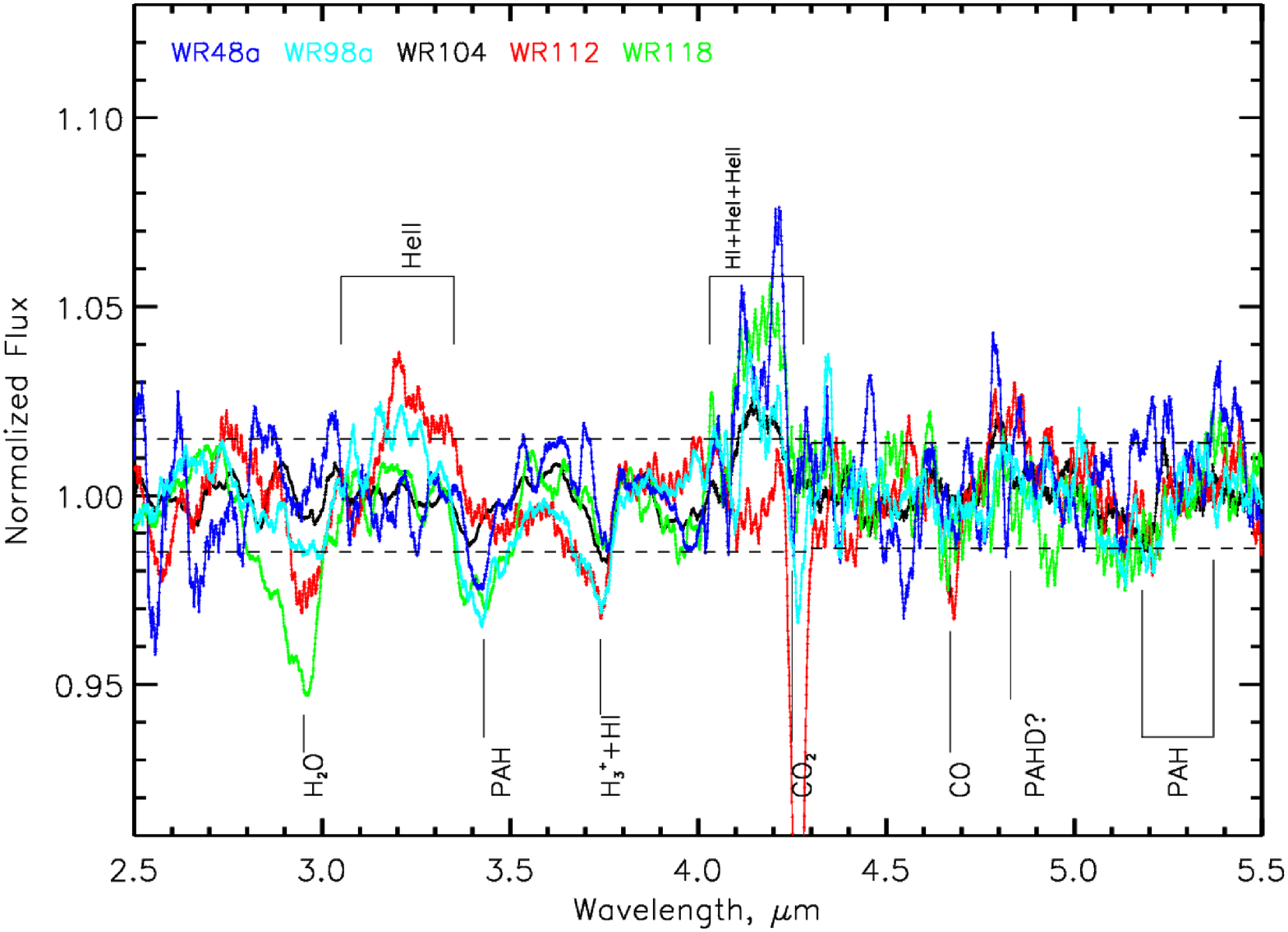}      
   \caption{Enlarged 2.5-5.5 $\micron\,$ region from Fig.~\ref{fig7}. }
    \label{fig8}
\end{figure}

\begin{figure}
 	\includegraphics[width=\columnwidth]{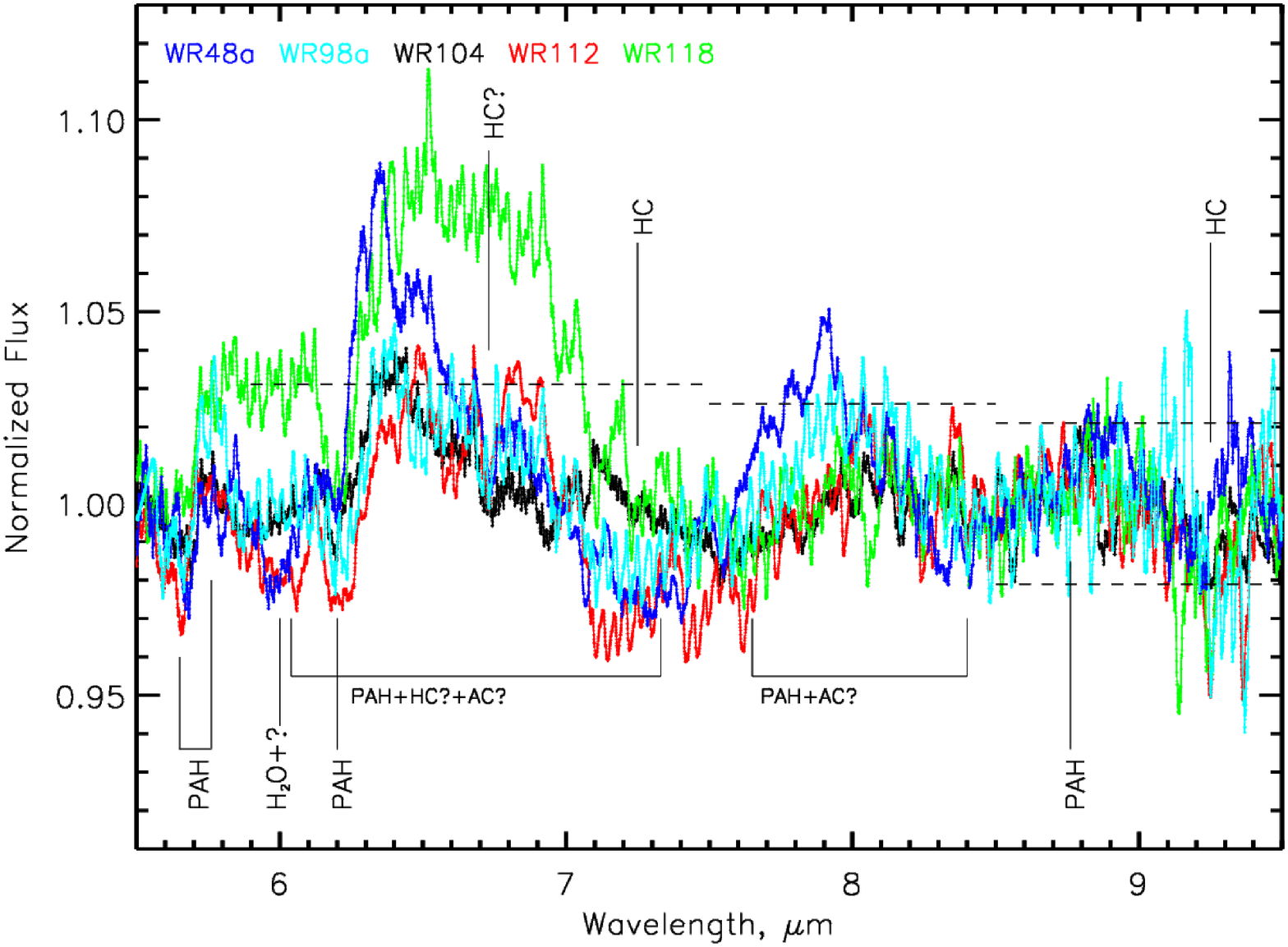}      
   \caption{Enlarged 5.5-9.5 $\micron\,$ region from Fig.~\ref{fig7}. }
    \label{fig9}
\end{figure}

\section{Discussion}

\subsection{Combined WR-PAH spectrum}

The combined WR-PAH spectrum (Fig.~\ref{fig1}) was assembled from the individual ISO spectra of five prodigious dust-producing WC stars.
All these stars belong to the special class of colliding-wind WR binaries (or candidate colliding-wind, since the optical spectra of these systems lack the features attributable to a companion star), known as dusty `pinwheel' nebulae (Tuthill et al. 1999, Monnier et al. 1999, Tuthill et al. 2006, Marchenko \& Moffat 
2007, Millour et al. 2009, Williams et al. 2012). In this binaries the stellar winds of massive binary companions (a Hydrogen-rich O-class star and a Hydrogen-deficient, however Carbon-rich WR star) meet and actively mix (Pittard 2007) along the wind-wind collision interface, thus injecting Hydrogen into the Carbon-rich WR environment and presumably forming PAHs along with copious amounts of Carbon dust. It is commonly believed that in such WR binaries the bulk of the dust is formed within the wind-collision zone (Williams 2008; Williams et al. 2009a), though for one of the sources used in this study, WR112, Lau et al. (2017) suggest that the extended `broken pinwheel' nebula emerged from a massive binary going through the red supergiant evolutionary stage, shortly before the WC phase. The spatially resolved infra-red images of these binaries bear close resemblance to `dusty pinwheels'. The freshly formed dust persists in the form of amorphous Carbon (e.g., Marchenko et al. 2002). 

\subsubsection{The IS (circumstellar?) absorptions}
 
The combined WR-PAH spectrum reveals a complex mix of absorption and emission details. The absorption features in the individual 
mid-IR spectra of the WR stars forming our WR-PAH composite were reported by many authors. Schutte et al. (1998) noticed the 3.0, 3.4, 4.3, 4.7, 6.0 and 6.2 $\micron\,$ absorptions in the ISO spectra of the `pinwheels', as well as a possible ~6.3 $\micron\,$ emission in WR104. Chiar et al. (2001) reconfirmed the presence of the 3.4 $\micron\,$ absorption in the spectra 
of all five `pinwheels' comprising our WR-PAH composite.
The distinctive 3.4 and 6.2 $\micron\,$ absorptions were seen in the spectrum of WR118 (Chiar \& Tielens 2001, Keane at al. 2001), as well as WR104 and WR112 (Chiar et al. 2001).  McCall et al (2002) noted the presence of the IS lines of H3$^+$ at 3.72 $\micron\,$ and CO at 4.65 $\micron\,$ in WR104, WR118 and WR121. Chiar et al. (2002a, 2011, 2013) followed up with detailed studies of the PAH and H$_2$O absorption features in the mid-IR spectra of massive stars. 
Beyond the known and well-studied IS and/or circumstellar absorptions in the spectra of Galactic massive stars (e.g., Chiar et al. 2002a, 2011, 2013), we note the presence of absorptions at 7.3 and 9.3 $\micron\,$ (Fig.~\ref{fig1}, Table 1). Therefore, the dust-related absorptions show a mix of features which belong to the high- (e.g., 3.4, 6.2, 6.8, 7.2 $\micron\,$) and low-temperature (3.3, 6.9, 9.3 $\micron\,$) formation pathways of carbonaceous grains (J\"ager et al. 2009). All five contributors to the WR-PAH spectrum
are heavily reddened objects, with $A_V \sim 7-14$ mag (van der Hucht, 2001)
usually attributed to absorption by a mix of IS and circumstellar dust. 
There is a clear correlation between the visual extinction and the optical 
depths of the IS 9.7 $\micron\,$ silicate absorptions, as well as the optical 
depths of the 3.4 $\micron\,$ PAH absorptions (Chiar \& Tielens 2001). 
This re-inforces the suggestion about the IS origin of the prominent 
$\lambda<$ 6.5 $\micron\,$ PAH absorptions in the WR-PAH composite.

\begin{table*}
\begin{minipage}{170mm}
	\centering
	\caption{The Identified PAH Features in the Combined WR-PAH Spectrum.}
	\label{tab:table_1}
	\begin{tabular}{llllll} 
		\hline
ID    &    $\lambda_{cent}$  &  EQW    & FWHM    &   Carrier\footnote{AC - amorphous Carbon; HC - hydrocarbons; PAHD - deuterated hydrocarbons; (?) - potential carrier; ? - unknown carrier; inter-stellar (IS) or circum-stellar (CS) or stellar-wind (SW) origin.}    &         References \\
  $\micron\,$   &    $\micron\,$      & $\micron\,$ & $\micron\,$    &         &   \\
		\hline
3.0 abs, IS    &   2.953  &  0.002  & 0.088     &     H$_2$0              &  Schutte et al. (1998), Chiar et al. (2000)  \\
3.4 abs, IS    &  3.402  &  0.002  & 0.079       &   PAH              &  Allamandola et al. (1989), van der Hucht et al. (1996) \\
3.5 abs, IS    &   3.481  &  0.001  & 0.072       &   PAH              &  Allamandola et al. (1989) \\
3.7 abs, IS    &   3.747  &  0.001  & 0.055       &   HI+H$_3^+$ (?)   &   McCall et al. (2002) \\
4.0 abs, SW    &  4.084  &  0.001  & 0.041     &    HI+HeI+HeII (?)    &   --                        \\      
4.0 em,  SW    &    4.172  &  0.003  & 0.128      &   HI+HeI+HeII     &   Lenorzer et al. (2002) \\
4.3 abs, IS    &   4.267  &  0.002  & 0.069      &   CO$_2$               & Schutte et al. (1998), Chiar et al. (2000) \\
4.7 abs, IS    &   4.673  &  0.001  & 0.047       &  CO                & Schutte et al. (1998), Pontoppidan et al. (2008)  \\           
4.8 em,  IS+CS? &   4.816  &  0.001  & 0.097      &   PAHD (?)          & Peeters et al. (2004), Buragohain et al. (2015) \\
5.2 abs,  IS    &   5.140  &  0.001  & 0.124      &   PAH                &  --                         \\
5.2 em,   CS?   &   5.400 &   0.001  & 0.144     &    PAH               & Peeters et al. (2002) \\
5.7 abs, IS     &   5.658 &   0.002  & 0.084      &   PAH               & van der Hucht et al. (1996)  \\                             
5.7 em,  CS?    &   5.745  &  0.002  & 0.127     &    PAH               & Peeters et al. (2002) \\
6.0 abs, IS     & $\sim$6.0  &   --    &    --   &       H$_2$O + ?        &    Schutte et al. (1998), Chiar et al. (2013)  \\
6.2 abs, IS     &    6.218  &  0.003  & 0.102       &  PAH               & van der Hucht et al. (1996), Chiar \& Tielens (2001), Chiar et al. (2013) \\
6.2 em, CS?        &    6.397  &  0.017  & 0.362     &    PAH               & Peeters et al. (2002), Chiar et al. (2002b) \\
6.8 abs, IS+CS?    &    6.736  &  0.001  & 0.056      &   HC (?)            & van der Hucht et al. (1996), J\"ager et al. (2009), Chiar et al. (2013) \\
6.8 em, CS?        &   6.825  &  0.012  & 0.267      &   HC + AC (?)       & J\"ager et al. (2009), Garc\'ia-Hern\'andez et al. (2013) \\
7.2 abs, IS+CS?    &    7.261  &  0.001  & 0.093     &    HC                & Sloan et al. (2007), J\"ager et al. (2009) \\
7.8 em, CS?        &    8.024  &  0.012  & 0.492     &    PAH + AC (?)      & Peeters et al. (2002), Chiar et al. (2002b), Garc\'ia-Hern\'andez et al. (2013) \\
8.6 em, CS?        &     8.760  &  0.008  & 0.704      &   PAH              &  Peeters et al. (2002) \\
9.3 abs, IS+CS?     &   9.246  &  0.001  & 0.046      &   HC                & J\"ager et al. (2009) \\
9.6 abs, IS?       &     9.612  &  0.001  & 0.085      &   ?                   &   --                  \\
10.0 abs, IS?      &   10.010  & 0.001  & 0.023      &   PAH$^+$ (?)        &   --                   \\
10.0 em, CS?       &   10.058 & 0.002  & 0.093      &   PAH$^+$ (?)      & Sloan et al. (1999)    \\ 
11.0 abs, IS?      &    10.981  & 0.003  & 0.071       &  ?                  &   --                   \\
11.3 abs, IS       &   11.134 &  0.002  & 0.076      &   ?                  &   --                    \\
11.3 em, IS+CS?    &    11.391  & 0.003  & 0.296       &  PAH             &  van Diedenhoven et al. (2004)  \\            
		\hline
	\end{tabular}
\end{minipage}
\end{table*}

\subsubsection{The WR-PAH emission line strengths} 

Now we turn to even more intriguing features in the WR-PAH composite, the broad emission details. Numerous studies point out that the C-C stretching modes produce 6.2 and 7.7 $\micron\,$ PAH bands, the C-H in-plane
deformation modes are responsible for the 8.6 $\micron\,$ emission, and the C-H out-of-plane deformation modes produce the 11.2 $\micron\,$ emission (L\'eger \& Puget 1984; Allamandola et al. 1985). The 11.3 $\mu$u PAH feature is very prominent (exceeding the strength of the 6.2 $\micron\,$ feature) in the integrated SEDs of the SMC and LMC (Israel et al., 2010; see also Melbourne \& Boyer 2013), as well as in star-forming galaxies (Smith et al. 2007). This band is almost invariably present in the circumstellar environments of young, massive Herbig Be stars (Verhoeff et al. 2012). The IS PAH spectra of highly reddened Galactic-Center WR stars (Fig.~\ref{fig1}) also show distinctive PAH features; however, note the apparent difference between the IS PAHs at 6.2 and 7.9 $\micron\,$ (Fig.~\ref{fig1}, lower section) and the presumably (see below) circumstellar WR-PAH emissions at $\sim$6.5 and 8.2 $\micron\,$ (Fig.~\ref{fig1}, upper section). 

Overall, the morphology of the WR-PAH composite spectrum is somewhat better matched by the general appearance of the generic type-B source emerging from some 
post-AGB stars, isolated Herbig AeBe stars and most PNe, with the PAH molecules formed in the stellar ejecta in predominantly pure-Carbon form (Peeters et al. 2002). This resemblance is strengthened by the relative weakness of the 11.2 $\micron\,$ emission in the class-B objects, commensurate with the surprisingly weak 11.2 $\micron\,$ PAH band in the combined WR-PAH spectrum (Fig.~\ref{fig1}). 
Such weakness may point to the prevalence of ionized PAH over neutral species in the WR environments. Indeed, the intense UV flux, as well as frequent dust-gas collisions (Zubko 1998) lead to a high degree of ionization of the WR wind and the embedded dust species. The ionized PAHs produce relativly enhanced, up to an order of magnitude in extreme cases, emissions in the $\lambda <$10 $\micron\,$ domain once compared to the 11.3 $\micron\,$ `benchmark' (Allamandola et al. 1999, Galliano et al. 2008b). 

\begin{figure}
	\includegraphics[width=\columnwidth]{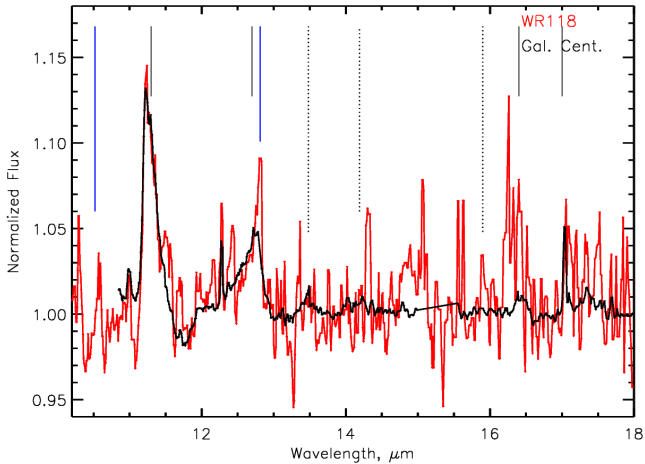}
    \caption{The normalized Spitzer WR118 spectrum (red line) and the Galactic Center ISM (black line), scaled to match the 11.2 $\micron\,$ feature of WR118. The assumed PAH positions are marked by black vertical lines; the [S IV] 10.52 and [Ne II] 12.81 $\micron\,$ emission lines are marked by blue vertical lines. The black vertical dotted lines mark the unidentified dust emissions from Smith et al. (2007).}
    \label{fig10}
\end{figure}

Modeling of the PAH spectra in the environments of young stars shows clear correlation of the $I_{6.2}/I_{11.3}$ ratio with the degree of PAH ionization (Maaskant et al. 2014). Comparing the strengths of the 11.3 $\micron\,$ spectral details in Fig.~\ref{fig1} and Fig.~\ref{fig10}, we conclude that practically all the 11.2 $\micron\,$ signal 
in the WR-PAH composite comes from the IS PAH feature observed in the individual WR118 spectrum (see below). Hence, such 11.2 $\micron\,$ emissions are likely absent in the spectra of the remaining 4 objects. The lack of any detectable 11.2 $\micron\,$ emission in the combined Gemini spectrum of WR48a and WR112 (Fig.~\ref{fig1}) attests to this assumption. The partial dehydrogenation of the PAHs (Allamandola et al. 1989) may also boost the strength of the $\lambda <$10 $\micron\,$ features, especially considering the wealth of energetic ($>$10eV) photons in the WR medium. This proves to be efficient for the small PAH clusters ($N_C<50$ Carbon atoms: Allain et al. 1996).  The well-defined IS 3.4 $\micron\,$ absorption in the WR-PAH composite, when considered together with the [presumably IS] 11.3 emission in the individual spectrum of WR118 (see below), points to the presence of neutral PAHs, most likely originating in the line-of-sight diffuse ISM. Hence, in the particular WR-PAH case we are not able to use the widely applicable PAH size diagnostics (Allamandola et al. 1989), since the 3.4 $\micron\,$ absorption and 11.3 $\micron\,$ emission likely belong to the IS PAH population, while we are primarily interested in the properties of the PAH species presumably arising from the WR circumstellar environments.   

To reproduce the overall appearance of the emission features in the WR-PAH spectrum, we use the on-line tools from the latest version of the NASA Ames PAH IR Spectroscopic Database (Bauschlicher et al. 2010; Boersma et al. 2014). We carefully remove the absorption features from the original
WR-PAH composite, subtract the continuum and bin the spectrum to 0.05 $\micron\,$ wavelength bins, limiting the fitting region to 4.3-12.0 $\micron\,$, in adherence to the recommended 5-15 $\micron\,$ range. 

\begin{figure}
	\includegraphics[width=\columnwidth]{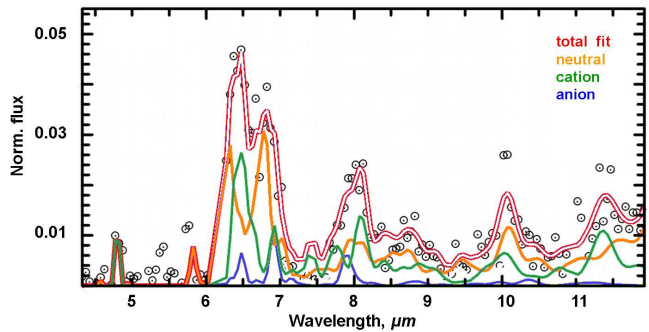}
    \caption{The model fits (lines of various colors representing different PAH species) to the cenzored (emissions-only) and wavelength-binned WR-PAH spectrum (open black circles).}
    \label{fig11}
\end{figure}

Experimenting with different combinations of PAH species, we find that incorporation of relatively large PAH clusters, $N_C > 50$, does not result in any major quality improvement of the fits: inclusion of PAH species with progressively higher amount of Carbon atoms (e.g., $N_C \le 30$, $N_C \le 40$, etc.) changes, at each step, the quality of the fit (measured by the $\chi^2$ statistics) by less than 1\%. On the other hand, the fits are rather sensitive to the choice of the lower bound of the PAH sizes, down to, but not beyond, $N_C <10$. Further lowering the size below this limit improves the quality of fit by less than 1\%. Hence, for the final fit (Fig.~\ref{fig11}) we chose the species with  $10\le N_C \le 30$, convolve the PAH cross-sections with a FWHM=25 $cm^{-1}$ (FWHM=0.16 $\micron\,$ at 8 $\micron\,$) Gaussian and apply the recommended redshft of 15 $cm^{-1}$, thus mimicking anharmonicity effects. In line with the general trends (Boersma et al. 2014), we note the relatively low contribution from anions and the large proportion of cations. We also note the surprizingly strong cation contribution to the 11.4 $\micron\,$ feature that exceeds the input from neutral species. As mentioned by 
Boersma et al. (2013), the PAH 5.2 and 5.7 $\micron\,$ features, as well as 6.8 $\micron\,$ aliphatic emission are not reproduced in the computed PAH spectra, hence note the misfit at 5.3-5.6 $\micron\,$ and the model-flux deficiency around $\sim$6.8 $\micron\,$. The dehydrogenated PAHs are also lacking in the current version of the database. Surprisingly, the unique (see below) 10.0 $\micron\,$ emission is also reproduced by a mix of neutral and ionized PAHs.        
The overall appearance of the fitted WR-PAH spectrum differs from the shapes of characteristic PAH spectra produced via stochastic mixing (Rosenberg et al. 2014) of the available theoretical PAH data, thus attesting to the unusual composition of the WR-PAH species.

He-burning WR stars have likely passed through a  Luminous Blue Variable (LBV) stage on their way to an imminent SN explosion. 
Hence, the conditions and chemistry of the LBV ejecta could serve as a proxy of the circumstellar environments of dust-producing WR stars. Both WR (Marchenko \& Moffat 2007) and LBV (Lau et al. 2014) show that, once formed, the dust survives for 10$^2$-10$^4$ years. However, compared to the WR-PAH appearance, the PAH bands detected in the shells ejected by LBV stars (or, alternatively, by their predesessors) show well-defined 11.2 $\micron\,$ features and distinctively different PAH intensity ratios in the $\lambda<10$ $\micron\,$ domain (cf. Voors et al. 1999, Umana et al. 2010, Niyogi et al. 2014).      
Another potential `proxy' also fails to match the peculiar appearance of the combined WR-PAH spectrum: the PAH features emerging from WR galaxies (i.e. galaxies with a sizable population of WR stars) 
show very prominent 6.2, 7.7 and 8.6 $\micron\,$ emissions (O'Halloran et al., 2005) and the 11.2 $\micron\,$ band which frequently exceeds
the strength of the 6.2 $\micron\,$ feature, contrary to the marginal presence of the former in the WR-PAH spectrum. 

Massive star-forming (starburst) regions (Brandl et al. 2012; Stock et al. 2014), AGNs (Shipley et al.l 2013), and young stellar objects (Oliveira et al. 2013) occupy distinctive places on the diagnostic $r_2=I11.3/I7.7$ vs $r_1=I6.2/I7.7$
diagram (intensity ratios of the corresponding PAH features: Draine \& Li 2001), with different types of PAH-emission objects 
usually clustering in $r_1<1$ (typically, 0.2-0.7) and $r_2<1$ (0.2-0.8) domain. The WR-PAH spectrum wanders far off the common boundaries, yielding 
$r_1\sim$2 and $r_2\sim1$, mainly dictated by the strength of the 6.5 $\micron\,$ complex and the relative weakness of the 7.7 $\micron\,$ WR-PAH emission. We also 
note that the surveyed (Galliano et al. 2008b) Galactic and Magellanic HII regions, and galaxies of various types (dwarf, spiral, starburst) cluster around $I6.2/I11.3 \sim1.0$, while the WR-PAH composite shows $I6.2/I11.3 \sim 2$. The broad and relatively intense 6.5 $\micron\,$ WR-PAH emission might be a complex blend (see below), hence heavily biasing the diagnostic 6.2/7.7 ratio towards unprecedentedly large numbers.

We evaluated the combined strength of the prominent 6.5 and 8.0 $\micron\,$ emissions in the individual spectra of the five WR stars 
contributing to the WR-PAH composite, then related these to the dust emissivity. Note that the latter completely dominates (Figs.~\ref{fig2}-6) the mid-IR flux for these dust-emitting stars. We find an average $\sim0.5\pm0.2$\% PAH flux contribution, broadly in line with the conclusion of Allamandola et al. (1985) that "... about 1\% of all the available Carbon is in the form of PAHs". However, under the assumption that the PAH and 
dust emissions come from the same circumstellar environments of the WR sources (essentially, from the WR gas-dust outflows), this WR-PAH fraction  
falls well below the more recent estimates: $\sim4.6$\% of the PAH fraction in the total dust mass of the Milky Way Galaxy (Draine and Li 2007), 3.9\% global PAH abundance for the Andromeda Galaxy (Draine et al. 2014). Star-forming galaxies show a broad distribution of the $L_{PAH}/L_{TIR}$
ratio (the integrated PAH and infrared luminosities) peaking around $\sim10$\% (Smith et al. 2007), thus also far exceeding the $<1$\% WR-PAH value. The majority of the modelled PAH/dust ratios in the galaxies with various heavy-element content fall in the $\sim1-20\%$ range, with only a handful of objects showing PAH/dust < 0.6\% (SBS 0335-052, Mrk 153, Tol 89, Mrk 930, II Zw 40: Galliano et al. 2008a). The low PAH/dust ratio could be linked to the very low Hydrogen content in the winds of single WC stars, which prevents 
the PAH nucleation along standard chemical pathways (Cherchneff 2010). On the other hand, in a massive colliding-wind binary, Hydrogen may be amply supplied by the wind of the O star companion, leading to PAH formation within the mixed-chemistry wind-wind interface (Cherchneff 2015). 
It can hardly be a coincidence that all the discovered WR-PAH sources are massive, colliding-wind (or candidate colliding-wind) binaries. However, such a wind-wind interface occupies only a small fraction of the circumstellar volume, thus severely limiting the efficiency of PAH formation. 

Besides the restrictive volume filling-factor of the wind-wind collision zone, the WR-PAH weakness could be ascribed to a combination of various additional factors. In particular, the low PAH strengths may also indicate rapid PAH processing (modification or destruction) in the presence of a strong ionizing radiation field (e.g., Gordon et al. 2008), especially considering the relative weakness of the 11.3 $\micron\,$ band and complete absence of the 12.7 $\micron\,$ feature in the combined WR-PAH spectrum. The WR-PAH composite apparently lacks the [S IV] emission at 10.52 $\micron\,$ that is usually indicative of an ionizing radiation field. This is not surprising, since such emissions are rarely detected in WC winds (Dessart et al. 2000). However, both the relatively weak 10.52 $\micron\,$ and much stronger [Ne II] 12.8 $\micron\,$ emissions are present in the spectrum of WR 118 (Fig.~\ref{fig10}), pointing to a strong ionizing field.  Such a circumstellar environment should lead to an accelerated modification/destruction of the material responsible for the PAH emissions, thus causing relative (to the underlying thermal emission from warm dust) weakness of the PAH bands (Gordon et al. 2008). Plus, one should also consider the destructiveness of the inherent and ubiquitous wind-embedded shocks (see Dessart and Owocki 2005 and references therein). The shocked, highly ionized environments of AGNs produce much weaker PAH features compared to the PAH spectra of their hosts (luminous infrared galaxies: Stierwalt et al. 2013). The observed EQW=0.017 $\micron\,$ of the WR-PAH 6.2 $\micron\,$ feature (Table 1) fits among the lowest values of the AGN-dominated group, defined as $EQW(6.2\,\mu m)_{AGN}\le0.1$ $\micron\,$ in Sargsyan et al. (2014), thus attesting to harshness of the dust-producing WR environment. However, one should note that PAHs were  unequivocally detected in an immediate vicinity of AGN (Alonso-Herrero et al. 2014).

\subsubsection{The WR-PAH redshifts}

The WR-PAH emission bands (Fig. 1) are apparently shifted in comparison to the emissions found in many categories of Galactic PAH sources (cf. Peeters et al. 2002). These systematic shifts cannot be related to any instrumental effects or artifacts of data processing during the compilation of the WR-PAH composite, since the well-defined IS absorptions show no apparent wavelength shifts exceeding $\sim0.02-0.05$ $\micron\,$. 
Among the known sources of PAH emissions we consider LBVs as the best evolutionary proxies to those of WR stars. Such consideration encompasses the average mass-loss rates, the luminosities and the temperatures of the central stellar sources, as well as physical conditions in the immediate circumstellar environments.
The PAH spectra of LBVs do not show any apparent redshifts beyond those attributable to the expected differences among various classes of PAH sources (e.g., Umana et al. 2010). 

On the other hand, a small but significant 0.1um red-shift was detected in the 11.2 $\micron\,$ PAH band arising from a giant HII region with prevalent photo-dissociative conditions surrounding a stellar association populated by early-type massive stars (Whelan et al. 2013).  
The physical conditions in a Galactic photo-dissociation region (an interface between UV source and a molecular cloud) may somewhat 
resemble conditions in the relatively remote, dust-generating regions of WC-star winds (e.g., Zubko 1998). In the surveyed photo-dissociation regions the $\sim0.1-0.3$ $\micron\,$ positional shifts in the 7.7-7.8 $\micron\,$ emission  
were attributed (Rapacioli et al. 2005; Bern\'e et al. 2007) to the radiation-induced evolution (photo-evaporation) of very small carbonaceous grains (longer-wavelength emission) into PAHs (shorter-wavelength spectral features). Such a scenario might be applicable to the WC case, where the models (Zubko 1998) predict an ample population of small Carbon grains. In addition, all the WC sources contributing to the WR-PAH composite are characterized by very high luminosities (on average, $log\, L/L_\odot\ga 5.2$ and high $T_{eff} >$ 40000K (Sander et al. 2012). 

\paragraph{The  6-7 $\micron\,$ emission complex} 

The broad, structured emission complex between 6-7 $\micron\,$ (Fig. 1) bears some resemblance to the `generic' 6.2 $\micron\,$ feature described by Peeters et al. (2002): the asymmetric (steep blue rise and extended red flank) emission, peaking at 6.2-6.3 $\micron\,$ is accompanied by a weak feature at 6.0 $\micron\,$. We interpret the apparent 
WR-PAH emission at $\sim6.1$ $\micron\,$ as arising from a combination of three spectral features: the broad IS absorption around $\sim6.0$ $\micron\,$, attributable to $\rm{H_2O}$ ice (Boogert et al. 2008), the IS PAH absorption at 6.2 $\micron\,$, and the broad WR-PAH emission at 6-7 $\micron\,$. We tentatively (note the question mark in Table 1) ascribe the $\sim6.0$ $\micron\,$ feature to $\rm{H_2O}$, however not being able to provide any meaningful measurements of its FWHM and EQW due to the ambiguity of defining a reference level, since this absorption is flanked by two emission peaks. Inspection of  
Figs.~\ref{fig8},~\ref{fig9} shows that both $\rm{H_2O}$ features, $3.0$ and $\sim6.0$ $\micron\,$, are clearly present in the ISO-SWS spectra of WR48a and WR112. Both features are absent in WR104, the star with the lowest $A_V$ among the considered objects. There is, however, the puzzling lack of  $\sim6.0$ $\micron\,$ feature in WR118 (the star with the highest $A_V$), despite the unequivocal presence of a strong $3.0$ $\micron\,$ absorption.

There is an additional IS absorption at 6.9 $\micron\,$ (e.g., Chiar et al. 2013) at the
top of the 6-7 $\micron\,$ complex. This complex peaks between 6.4 and 6.5 $\micron\,$, well beyond the 6.2-6.3 $\micron\,$ range observed in the various classes of  Galactic PAH sources.  
The redshift, strength, FWHM and extension of the red flank of the 6-7 $\micron\,$ WR-PAH emission are unprecedented: cf. the average $FWHM\le0.15$ $\micron\,$ for the PAH classes A-C to $FWHM\sim0.5$ $\micron\,$ for the 6-7 $\micron\,$ WR-PAH complex. Such a redshifted 6-7 $\micron\,$ complex could arise 
from a population of small ($N_C \la20$ atoms: cf. the dependence of the redshift on the PAH size in Peeters et al. 2002), ionized PAHs, presumably formed in the winds of massive dust-producing WC+O binaries. In addition, the redshift could be caused by the charge (Bakes et al. 2001) and dehydrogenation (Pauzat et al. 1997) of the PAH molecules, with the latter cause quite applicable to the WC environment. Indeed, the modeled WR-PAH spectrum (Fig.~\ref{fig11}) definitively points to the presence of ionized, relatively small ($N_C \la30$) PAH species.

However, there is an alternative explanation of the 
broad 6-7 $\micron\,$ feature, promoted by the recent discovery of the emissions from carbonaceous grains in the outflows from the extremely Hydrogen-deficient 
R Coronae Borealis (RCB) stars. A comprehensive survey of the Galactic RCB spectra revealed the presence of emission features at   
5.9, 6.3, 6.9, 7.3, 7.7, 8.1, 8.6, 9.1, and 9.6 $\micron\,$ (Garc\'ia-Hern\'andez et al. 2013). The chemistry of RCB outflows closely resembles the Hydrogen-deficient, Carbon-rich winds of WC stars.  Hence, it is very tempting to interpret the broad, red-shifted 
WR-PAH features as emerging from a blend of PAH and amorphous-Carbon emissions, especially noting that Hydrogen-poor environments tend to produce relatively red-shifted (by as much as $\sim0.2$ $\micron\,$) emissions at 6.4 $\micron\,$ (Garc\'ia-Hern\'andez et al. 2013). Such an interpretation may specifically apply to the ~6.5 $\micron\,$ WR-PAH complex. 
The apparent split of the line-feature peak into (at least) two components with emissivity maxima around 6.4 and 6.8 $\micron\,$ (Fig.~\ref{fig1}) suggestively matches the 6.3 and 6.9 $\micron\,$ features from the RCB list. Then, the broad 8.0 $\micron\,$ WR-PAH feature could be ascribed to the dominating 8.1 $\micron\,$ band in the RCB spectra. There is also noticeable weakness or complete absence of the 11.2 and 12.7 $\micron\,$ PAH features in the RCB spectra (Garc\'ia-Hern\'andez et al. 2013), in line with the general appearance/absence of these features in the WR-PAH composite spectrum. 

\paragraph{The $\sim$8.0 $\micron\,$ emission} 
 
The second strong WR-PAH emission peaks around $\sim8$ $\micron\,$, i.e. slightly redward of the characteristic $7.6-7.8$ $\micron\,$ emission observed in the majority of PAH sources. This $\sim0.1-0.2$ $\micron\,$ redshift seems to be less dramatic 
compared to that of the 6-7 $\micron\,$ complex. Some (though relatively few) Galactic B$\arcsec$ and C$\arcsec$ PAH sources (as classified in Peeters et al. 2002) show peaks at $\lambda > 7.9$ $\micron\,$. 

\paragraph{The $\sim$8.8 $\micron\,$ emission} 

The weak, symmetric WR-PAH emission at $\sim8.8$ $\micron\,$ also reveals a $\sim0.15$ $\micron\,$ redshift compared to the generic A and B PAH sources. This shift cannot be linked to the presence of the putative emission features from carbonaceous grains. We already mentioned that ionized PAHs may modify the general appearance of the WR-PAH composite spectrum. The theoretical spectra of compact, symmetric, ionized PAHs with $N_C \ga 100$ show pronounced ($\ga 0.2$  $\micron\,$) redshifts for the 6.5 $\micron\,$ and 8.0 $\micron\,$ bands (Ricca et al. 2012). The $\sim8.6$ $\micron\,$ emission remains relatively unperturbed, until $N_C$ exceeds 200, thus resulting in a $\sim0.3$ $\micron\,$ redshift. However, a model fit of the (presumably circumstellar) WR-PAH spectrum (Fig.~\ref{fig11}) shows prevalence of the $N_C < 50$ species in the $\lambda< 8$ $\micron\,$ domain. With the increasing impact of large PAH agglomerates at longer wavelengths, the observed moderate red-shift of the $\sim8.8$ $\micron\,$ emission may come from a mix of the (large) IS and (relatively small) circumstellar PAH clusters.          

\subsubsection{The $\sim$10.0 $\micron\,$ emission}

The combined WR48a and WR112 spectrum obtained with Gemini (Fig.~\ref{fig1}) shows a relatively narrow emission at 10 $\micron\,$, far exceeding the 
detection threshold. This emission also shows up, though much less prominently, in the combined WR-PAH spectrum produced from the ISO-SWS data, with WR48a (Fig.~\ref{fig2}; readily visible) and WR112 (Fig.~\ref{fig5}; seemingly hidden in the noise) as equal contributors among 5 sources. 
Reports on similar spectral features are scarce.  
The reflection nebula NGC 7023 illuminated by a luminous, massive early-type  star, shows prominent PAH features along with the cationic buckminsterfullerene $C_{60}^+$ (Berne et al. 2013). This case provides an example of the interstellar environment with a strong UV field somewhat resembling the surroundings of a late-type WC star.
Besides the presence of a well-defined 11.3 band, the NGC 7023 region also shows a very faint 10.25 $\micron\,$ emission feature (cf. Fig. 1 and Fig 2 in Berne et al. 2013). 
A very weak, broad emission blend around $\sim$10 $\micron\,$ also shows up in the continuum-subtracted spectrum of the LBV star HD 168625 (Umana et al. 2010). The most unequivocal detection of a broad $\sim10$ $\mu$ emission feature comes from space-resolved spectroscopy of the region in the reflection nebula NGC 1333 (Sloan et al. 1999) where the strength of the 9.8 $\micron\,$ feature strongly depends on the distance to the central source. The frequently detected $\rm{H_2}$ S(3) emission (e.g., Stierwalt et al. 2014) peaks around 9.7 $\micron\,$, thus with certainty not being able to contribute to the WR-PAH emission.    

The carrier of this emission in the combined spectrum of WR48a and WR112 remains elusive. We note the lack of any distinguishable spectral feature around 11.2 $\micron\,$ in the Gemini spectrum (the blue dashed line in Fig.~\ref{fig1}). Otherwise, the combination of 10 and 11.2 $\micron\,$ emissions would have pointed to the presence of olivine (e.g., Aller et al. 2012). Hence, we side with Sloan et al. (1999), ascribing the 10 $\micron\,$ emission to ionized PAH cations.  

The `kernel' PAH spectra (Rosenberg et al. 2014) produced from the NASA Ames PAH IR Spectroscopic Database (Bauschlicher et al. 2010; Boersma et al. 2014) lack any traces of this prominent 10 $\micron\,$ emission. However, in our analysis this feature could be partially reproduced (Fig.~\ref{fig11}) by an $\sim$equal mix of ionized and neutral PAHs with $N_C \le30$. 

\subsection{The individual PAH sources: WR 19 and WR 118}

Besides the ISO-SWS spectra of the `dusty WR pinwheels', we retrieved and analyzed all available archival Spitzer spectra of the known dust-producing WR stars WR19, WR53, WR48a, WR76, WR98a, WR103, WR118, and WR140. We found a clear presence of PAH emissions in WR19 and WR118. Distinguishing between the strong wind emissions and PAH features, we used the high-quality spectra from the Spitzer atlas (Ardila et al. 2010). Additionally, we also analyzed the spectra of the heavily reddened Galactic Center WR stars of type WN (none known to be dust producers): WR102a, WR102b, WR102c, 
WR102j, and WR102ka. We found that the Galactic-center WR population shows clear presence of PAHs practically indistinguishable form the IS PAH features (Fig.~\ref{fig1}). E.g., the strong 11.2 $\micron\,$ PAH emission in WR102ka is most likely coming from the surrounding LBV-like nebula (Barniske et al. 2006).  
    
\begin{figure}
	\includegraphics[width=\columnwidth]{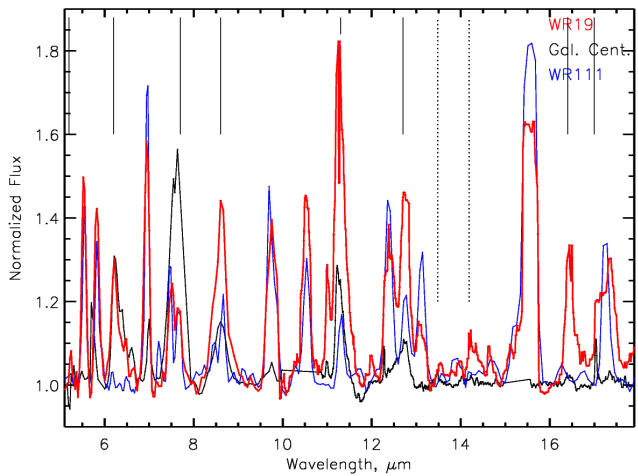}
    \caption{The normalized Spitzer WR19 spectrum (red line), the scaled (to provide the best match to the WR19 lines in the $\lambda < 8$ $\micron\,$ region) spectrum  of a single, dust-free star WR111 (blue line) and the Galactic-Center PAH specrum (black line), scaled to match the 6.2 $\micron\,$ feature in the WR19 spectrum. The assumed PAH positions are marked by black vertical lines. The black vertical dotted lines mark the unidentified dust emissions from Smith et al. (2007).}
    \label{fig12}
\end{figure}

The colliding-wind binary WR19 is known for periodic outbursts of dust production (Williams et al. 2009b). The Spitzer spectrum of WR19  shows exceptionally strong 11.3, 12.7, 16.4 $\micron\,$ PAH features, with a hint of a weak 17.0 $\micron\,$ complex. However, it lacks the 7.7 $\micron\,$ PAH emission all together
(Fig.~\ref{fig12}). The absence of the 7.7 $\micron\,$ PAH band is somewhat surprising, considering the reasonably well-correlated behavior of mid-IR PAH emissions in 
practically all classes of PAH sources (e.g., Whelan et al. 2013). Such an absence was noticed in a very limited sample of objects, initially in two post-AGB stars    
(Peeters et al. 2002). Then it was seen in a group of objects where the PAH molecules are formed either in a circumstellar disk or an outflow excited by a relatively cool ($T_{eff} < 10000$ K) star with a dearth of ionizing UV flux, leaving the PAH material practically unprocessed (Sloan et al. 2007, Gielen et al. 2009 and references therein). The case of WR 19 brings some challenges to such an interpretation, as $T_{eff}$ of the WC5 companion in this binary may exceed 80 000 K, judging by the average $T_{eff}=83 000$ K for this spectral subclass (Sander et al. 2012). On the other hand, in WR19 the general appearance of the PAH spectrum 
resembles the morphology of a C-type PAH source (e.g., Gielen et al. 2009), where, besides the lack of the 7.7 $\micron\,$ spectral feature, the strength of the 11.3 $\micron\,$ emission exceeds that at 6.2 $\micron\,$. Thus, such a PAH spectrum may arise from an intervening, but distant background source (either extended or line-of-sight stellar).    

The mid-IR ISO spectrum of the dust-producing star WR118 (a potential `pinwheel' binary: Millour et al. 2009) contributed to the WR-PAH composite.  
The Spitzer spectrum of WR118 (Fig.~\ref{fig10}) shows distinctive 11.2, 12.8 $\micron\,$ PAH features and a surprisingly strong 16.4 $\micron\,$ band, without any noticeable red-shifts in the 11.2 and 12.8 $\micron\,$ emissions. This points to a potential presence of very large, neutral (Schutte et al. 1993; Hudgins \& Allamandola 1999; Draine \& Li 2007), presumably interstellar PAHs 
in the line of sights towards WR19 and WR118. On the other hand, the 11.2 $\micron\,$ feature only marginally appears in the combined WR-PAH 
spectrum (with WR118 as an equal contributor among the five sources) which, at the same time, shows the red-shifted, peculiar WR-PAH profiles in the $\lambda<$10 $\micron\,$ domain. This may signify that we see a mix of the two PAH populations:
(1) the large, neutral inter-/circum-stellar (swept up?) PAH grains in WR19 and WR118, and (2) the smaller, probably ionized PAHs as seen in the combined WR-PAH spectrum. The presence of [SIV]
10.52 $\micron\,$ and [NeII] 12.81 $\micron\,$ emissions is usually linked to strong, ionizing radiation fields.  [NeII] and (possibly) [SIV] are detectable as weak features in the WR 118 spectrum (Fig.~\ref{fig10}), which probably means that they originate from a rather distant circumstellar shell. We also note the presence of an unidentified $\sim14.2$ $\micron\,$ emission in the 
spectra of WR19 and WR118. This weak emission, designated as `dust feature' by Smith et al. (2007) is also seen in the spectra of star-forming galaxies. 
 
\subsection{Immediate surroundings of the dust-producing WR-PAH sources}

The mid-IR spectra of heavily reddened Galactic-Center WR stars show clear presence of the PAH bands (Fig.~\ref{fig1}). However, it appears that for these particular stars the PAH features arise from the interstellar medium, since  WN stars (WN8 for WR102a and WN6 for WR102b) are not regarded as credible dust-producing sources. The locations and shapes of the presumably interstellar Galactic-Center PAH features match the IS PAH spectrum excited by the UV radiation fields of the hot and luminous Orion-belt stars (Fig.~\ref{fig1}, lower panel). The combined WR-PAH spectrum (Fig.~\ref{fig1}, upper panel) shows distinctively different morphology. However, this could be regarded only as a relatively circumstantial argument in favor of the different origin of these features, when taking into consderation the extremely wide variety of the observed PAH shapes. More arguments may come from inspection of the immediate surroundings of the WR stars contributing to the WR-PAH composite. 

\subsubsection{WR48a}
The high-resolution 12.3 $\micron\,$  image of WR48a (Marchenko \& Moffat 2007) shows an extended ($\sim10\arcsec$) dust emission with concentric, asymmetric arcs and distinctive blobs. These relatively faint circumstellar structures are not resolved in the WISE 3.4, 4.6, 12 and 22 $\micron\,$ images. The Spitzer/IRAC mid-IR maps show strong, extended IR emission bordering WR48a. However, the extended source is fairly ($\sim100\arcsec$) distant from the WR48a system, thus not likely to contribute to the analysed ISO-SWS spectra that were taken with $14\arcsec\times20\arcsec$ aperture (de Graauw et al. 1996).  The dust emission seen at 2.2 $\micron\,$  predominantly (85\%) comes from a very compact central source with $FWHM=0.07\arcsec$ (Monnier et al. 2007). The $K_s$ 2MASS image shows an apparently single source in a crowded field populated by much fainter objects. 

\subsubsection{WR98a}
The dust spiral around WR98a was not seen (i.e., size of the `pinwheel' was below the $\sim0.5\arcsec$ resolution limit) in the mid-IR (8-18 $\micron\,$) images obtained by Marchenko \& Moffat (2007). The 3.1 $\micron\,$ images from Monnier et al. (2007) show an extended, however relatively compact (characteristic size $0.14\arcsec$) source that traces the `pinwheel' shape clearly observable in the contemporaneous 2.2 $\micron\,$ image. The $K_s$ 2MASS image shows a relatively distant ($\sim30\arcsec$, thus not likely entering the ISO-SWS aperture), fainter companion. WISE images depict relatively close (presumably stellar) sources that may contrinute to the $<5$ $\micron\,$ flux. Indeed, Wallace et al. (2012) resolve the system into a compact group of 4 stars in the V,R,I Hubble Space Telescope (HST) images. However, WR98a clearly dominates the field at longer ($\lambda>3$ $\micron\,$) wavelengths. The WISE and Spitzer/IRAC maps lack any extended mid-IR sources in the immediate ($<50\arcsec$) surroundings of WR98a that can make significant contributions to the ISO-SWS flux. 

\subsubsection{WR112}
The 3.8 $\micron\,$  images of WR112 obtained by Marchenko et al. (2002) show a clearly extended source. It's shape conforms, if roughly, to the innermost emission distribution at 8 $\micron\,$. Monnier et al. (2007) also finds an elongated, though compact (mean $FWHM=0.13\arcsec$) source at 3.1 $\micron\,$.  The 12.3 $\micron\,$ image from Marchenko \& Moffat (2007) shows a 'broken pinwheel' nebula with at least 5 regularly separated arcs spread over $\sim10\arcsec$. Despite this large extension, at least 90\% of the measured mid-IR flux comes from the central $3\arcsec$ part of the dust nebula (Lau et al. 2017). Hence, the source appears point-like and isolated in the relatively low-resolution WISE and Spitzer/IRAC images, despite the presence of a close ($\sim1\arcsec$: Wallace et al. 2012) companion in the broadband U,B,V images obtained by the HST Wide Field and Planetary Camera 2 (WFPC2).

\subsubsection{WR104 and WR118}
Marchenko et al. (2002) found only modestly ($^<_\sim0.5\arcsec$) extended 8-12 $\micron\,$  dust emission from WR104 and WR118, in general agreement with esimates from Monnier at al. (2007): both sorces seen as barely extended at 3.1 $\micron\,$, with  $FWHM=0.1\arcsec$ (WR104) and $0.03\arcsec$ (WR118). Both WR104 and WR118 are embedded into IR-bright surroundings. High-resolution WFPC2/HST images resolve WR104 into a double source with $\sim1\arcsec$ separation (Wallace et al. 2012). WR104 appears to be single in the lower-resolution $K_s$ 2MASS images that show no close comparable-brightness companions to within $\sim20\arcsec$. 
WR104 completely dominates the field in mid-IR images from WISE and Spitzer/IRAC. Some faint filaments seen in the WISE 12 $\micron\,$ images are sufficiently distant for not contributing to the ISO-SWS spectrum. 
On the other hand, the Spitzer/IRAC mid-IR images show WR118  embedded into a faint diffuse emission. This is clearly seen at 12 and 22 $\micron\,$ (WISE data). Hence, the sufficiently close filaments may add to the $\lambda>10$ $\micron\,$ flux. 
Besides, the $K_s$ 2MASS field shows WR118 as an isolated source without any companions of comparable brightness that may significantly contribute to the mid-IR flux.  

\subsubsection{WR19}
As we already mentioned, the morphology of the WR19 PAH spectrum points to its likely origing in the IS environment. WR19 lies at the outskirts of a bright IR region (clearly seen in the WISE and Spitzer/IRAC survey data) that possibly contributes to the $\lambda>10$ flux. Moreover, there is a bright, close line-of-sight object seen in the 3.4 and 4.6 $\micron\,$ WISE images. Both these line-of-sight sources may contribute to the WR19 PAH spectrum.

Comprehensive surveys of nebulae around Galactic WR stars find that up to $\sim35$\% of them are surrounded by large shells seen both in the optical and IR images (Marston 1996, 1997). Some cold, massive circumstellar shells carry traces of molecular gas: HCN, HCO$^+$, CN, and HNC was found by Marston (2001) in the Wolf-Rayet ring nebula NGC 3199 around WR18.  
Such cold, massive circumstellar shells may also carry PAH complexes synthesized in a wind of WR predecessor. However, none among the five WR stars contributing to the WR-PAH spectrum is known as posessing any detectable circumstellar ring-like structure. Inspection of the available images show  WR48a, 98a, 104 and 112 as sufficiently isolated, bright mid-IR sources, with a domineering (typically, $\ga90$\%) emission coming from central parts.  Hence, in the  $\lambda\la$10 $\micron\,$ domain the average WR-PAH spectrum most likely originates from the immediate surroundings of these massive WR binaries, with the PAH species presumably produced in the colliding-wind wakes.   
 
\subsection{PAH in the early Universe}

Both dust and PAH species show surprising resilience in the harsh UV environments ranging from the centers of young, massive stellar clusters to AGN nuclei (Alonso-Herrero et al. 2014; Stolte et al. 2015).  
The characteristic 0.2175 $\micron\,$ feature is commonly linked to IS absorption by PAH and graphite. It was detected (Scoville et al. 2015) in the spectra of high-redshift, z=2-6.5, galaxies, potentially pointing to an early-epoch accumulation of PAH species. The 6.2 $\micron\,$ PAH emission was directly observed in a z=4 galaxy (Riechers et al. 2014), and a copious amount of primordial (z=7.5) dust was detected in a star-forming galaxy, A1689-zD1 (Watson et al. 2015).  In the modern Universe the WR stars are viewed as relatively insignificant (in comparison to SN ejecta and AGB outflows) dust sources. However, the WR dust output may rival the AGB yield at $z>10$, judged by the models from Dwek \& Cherchneff (2011).  Hence, over the first $\sim$0.5B years, the primordial PAH species could be produced in the winds of the rapidly evolving WR stars, considering the following. The injection into ISM of the PAH species produced in the AGB outflows is delayed (Galliano et al. 2008a), mostly due to the relatively long evolutionary timescales for low-mass stars. Thus, the core-collapse supernovae are viewed as domineering sources of dust in the very high-redsift Universe (Dwek et al. 2014), with WR stars contributing additional $\sim 10$\% (Galliano et al. 2008a). However, despite numerous observations of dust production in the SN outflows, there is a lack of evidence that PAH may be generated in the SN environments. Tappe et al. (2012) finds that the circumstellar PAH are rather swept up and eventually destroyed by a SN blast wave. This leaves WR stars as potentially unique PAH sources in the early Universe.    

\section{Conclusions}

The combined 3-12 $\micron\,$ spectrum of colliding-wind, dust-generating Wolf-Rayet binaries WR48a,98a,104,112 and 118 shows a wealth of emission and absorption features. 
We link the former to relatively small-size (probably, $N_C \ll 100$ atoms) circumstellar PAH, while the latter represents the extensively studied IS PAH population. The assumption about the relatively small size of the (circumstellar) WR-PAH features is backed by two agruments: by the general appearance of the WR-PAH spectrum
(strong, well-developed $\sim$6.5 $\micron\,$ emission complex in contrast to the unusually faint 11.2 $\micron\,$ PAH feature), as well as by the theoretical fits to the WR-PAH spectrum (Fig.~\ref{fig11}) favoring species with $10 \le N_C \le 30$.    

The overall weakness of the detected emission features, 11.2 $\micron\,$ in particular, points to PAH reprocessing in the presence of strong ionizing fields and ubiquitous wind-embedded shocks. We tentatively link the 10.0 $\micron\,$ emission to a population of positively-charged PAHs.  

Though we are not able to unequivocally conclude that the PAH features are formed in the dusty envelopes of the binaries, 
the unique appearance of the PAH spectrum and especially the substantial, $\ga0.2$ $\micron\,$, systematic red-shifts of the PAH features support such a suggestion. Moreover, four out of five (with WR118 as a possible exception, however only in the $\lambda>$10 $\micron\,$ domain) sources contributing to the combined WR-PAH spectrum appear to be relatively compact and sufficiently isolated from any IR foregrounds/backgrounds (both extended and point-like) of a comparable brightness, thus increasing the probability that the weak PAH emissions are emerging from the outflows of these massive colliding-wind binaries. 

The long-wavelength, $\lambda>$10 $\micron\,$ PAH emissions in the spectra of the colliding-wind binaries WR 19 and WR 118 may arise from the diffuse circumstellar or interstellar environment, pointing to the presence of large, neutral PAH carriers. 
Indeed, the 12 and 22 $\micron\,$ WISE images of WR118 show the presence of bright filaments in the immediate surroundings of the star. We also note the peculiar absence of the 7.7 $\micron\,$ PAH feature in the PAH spectrum of WR19.

\section*{Acknowledgements}
The Spitzer Space Telescope is operated by the Jet Propulsion
Laboratory, California Institute of Technology, under NASA
contract 1407.  This research has made use of the NASA/IPAC
Infrared Science Archive, which is operated by the Jet Propulsion Laboratory, 
California Institute of Technology, under contract with the National Aeronautics Space Administration,
as well as the NASA Ames PAH IR Spectroscopic Database.
Based [in part] on observations obtained at the Gemini Observatory, which is operated by the Association of Universities for Research in Astronomy, Inc., under a cooperative agreement with the NSF on behalf of the Gemini partnership: the National Science Foundation (United States), the National Research Council (Canada), CONICYT (Chile), Ministerio de Ciencia, Tecnolog\'{i}a e Innovaci\'{o}n Productiva (Argentina), and Minist\'{e}rio da Ci\^{e}ncia, Tecnologia e Inova\c{c}\~{a}o (Brazil). AFJM is grateful to NSERC (Canada) and FQRNT (Qu\'ebec) for financial assistance. We thank P. Williams and J. Chiar for numerous helpful comments and suggestions.


\bsp	
\label{lastpage}
\end{document}